\documentclass[final,preprint,tightenlines,floatfix,nofootinbib,superscriptaddress,fleqn]{revtex4}
\usepackage{graphicx}
\usepackage{bm}
\usepackage{dcolumn}
\usepackage{amsmath}%
\usepackage{amsfonts}%
\usepackage{amssymb}

\newcommand{\beq}{\begin{equation}}
\newcommand{\eeq}{\end{equation}}
\newcommand\beqa{\begin{eqnarray}}
\newcommand\eeqa{\end{eqnarray}}
\newcommand\bea{\begin{array}}
\newcommand\eea{\end{array}}

\newcommand{\neqa}{\nonumber\end{eqnarray}}
\newcommand{\la}{\label}

\newcommand{\CO}{{\ensuremath \mathcal O}}

\def\Dirac#1{#1\hskip-6pt/}
\def\dd{\Dirac\partial}

\def\mpcomm#1{\nextline\strut\kern-6em{\tt MP COMMENT => \ #1}\nextline}
\def\nextline{\hfill\break}

\begin{document}

\preprint{PNU-NTG-10/2004}

\title{Pentaquarks: Review on Models and Solitonic Calculations
  of Antidecuplet Magnetic Moments}
\author{Klaus Goeke}
\email{Klaus.Goeke@tp2.ruhr-uni-bochum.de}
\affiliation{Institut f\"ur Theoretische Physik II,
Ruhr-Universit\" at Bochum, D--44780 Bochum, Germany}
\author{Hyun-Chul Kim}
\email{hchkim@pusan.ac.kr}
\affiliation{Department of Physics, and Nuclear Physics \&
Radiation Technology Institute (NuRI), Pusan National University,
609-735 Busan, Republic of Korea}
\author{ Micha{\l} Prasza{\l}owicz$^{c}$}
\email{michal@th.if.uj.edu.pl}
\affiliation{M. Smoluchowski Institute of Physics,
Jagellonian University, ul. Reymonta 4, 30-059 Krak{\'o}w, Poland}
\author{Ghil-Seok Yang}
\email{gsyang@pusan.ac.kr}
\affiliation{Department of Physics, and Nuclear Physics \&
Radiation Technology Institute (NuRI), Pusan National University,
609-735 Busan, Republic of Korea}

\begin{abstract}
~~~\newline After a quick look at the present status of
experiments we shortly review in a first step the ongoing
discussions on some of the basic theoretical approaches to
pentaquarks: Chiral solitons, Large-$N_c$ considerations,  quark
models, and lattice gauge approaches. In a second step we apply
the chiral quark-soliton approach including linear
$m_s$-corrections to the calculation of magnetic moments of the
baryons of the anti-decuplet. In this approach the parameters are
fixed solely by experimental data for the magnetic moments of the
baryon octet and for the masses of the octet, decuplet and of
$\Theta^{+}$. The magnetic moment of $\Theta^{+}$ depends rather
strongly on the pion-nucleon sigma term and reads $-1.19\,{\rm
n.m.}$ to $-0.33\,{\rm n.m.}$ for $\Sigma_{\pi N} = 45$ and $75$
MeV respectively. As a byproduct the strange magnetic moment of
the nucleon is obtained with a value of $\mu^{(s)}_N =+0.39$ n.m.
\end{abstract}
\maketitle

\section{Introduction}

\begin{figure}[t]
\begin{center}
\includegraphics[scale=1.1]{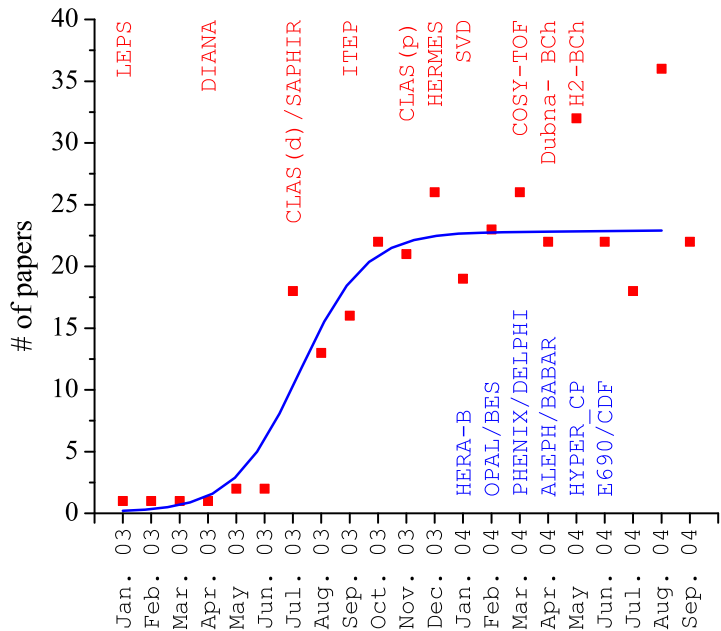}
 \caption{\it
      Evolution of the total number of manuscripts discussing
      pentaquarks during the last months and experiments reporting  the
      observation (top) or non-observation (bottom) of the $\Theta^+(1530)$ signal.
      The solid line represents an eye-guided fit.
    \label{fig:WA8901} }
\end{center}
\end{figure}

The narrow baryonic resonance $\Theta^{+}$ with mass of $1530$~MeV
and strangeness of $+1$ (i.e. containing one excess strange
anti-quark) and width of about $\Gamma < 15$~MeV has last year
been identified by several groups
~\cite{Nakano:2003qx,experiments}. Such a state is extremely
exciting because it is unambiguously exotic in the sense that it
cannot be a simple three-quark state. These experiments have been
triggered by predictions of position and width in the chiral quark
soliton model ($\chi$QSM) by Diakonov, Petrov and Polyakov
\cite{Diakonov:1997mm}. An earlier estimate of the mass in the
soliton approach of the Skyrme model was given by Prasza{\l}owicz
\cite{Praszalowicz:2003ik}. The discovery of $\Theta^+$ together
with the accurate prediction of Ref.\cite{Diakonov:1997mm}
 have initiated considerable theoretical
activity yielding more than 200 papers on this subject. The
discussion got even very heated since in the last months several
other experiments did not show any evidence of the $\Theta^{+}$.
Actually no signals of the $\Theta^+$ have been found by a number
of groups
\cite{Theta:BES,Xi:HERAB,Theta:OPAL,Theta:PHENIX,Theta:DELPHI,
Theta:ALEPH,Theta:E690,Theta:CDF,Theta:BABAR}. See Figure
\ref{fig:WA8901} based on Ref.~\cite{Pochodzalla}. Besides the
$\Theta^+$ there is perhaps an observation of an exotic
$\Xi_{\overline {10}(1860)}$ state by the NA49 experiment at
CERN~\cite{Alt:2003vb}, though it is still under debate. Much of
the theoretical activity has been aimed at understanding the
structure of these exotic states, $\Theta^+$ in particular.
Besides using the chiral solitonic approach the most common
treatment of this problem has been based on variants of the quark
model where the new baryon is identified as a
pentaquark~\cite{JW,qm}. Other approaches treat the $\Theta^+$ in
terms of  meson-baryon binding~\cite{IKOR,KK} or as a
kaon-pion-nucleon state~\cite{LOM}. While these approaches are all
interesting, they are also all highly model-dependent and it is
difficult to assess in an {\it a priori} way their validity. A
discussion of some of the issues raised by various models may be
found in \cite{JM,Jennings:2003wz,Zhu:2004xa}. Attempts to
identify quantum numbers of $\Theta^+$ and to confirm/deny its
existence directly from QCD have been performed within the lattice
Monte-Carlo simulations
\cite{Fodor,Csikor:2003ng,Sasaki:2003gi,Liu,Chiu:2004gg,Chiu:2004uh,Sasaki:2004vz,newlattice}
and within the QCD sum-rules \cite{sumrules}.

The analysis based on the SU(3) chiral soliton model treated with
rigid-rotor
quantization~\cite{Praszalowicz:2003ik,Diakonov:1997mm} appears
different from other treatments of the $\Theta^+$ structure in a
number of ways. In fact, exotic SU(3) representations containing
exotic baryonic states are naturally accommodated within the
chiral soliton models~\cite{Chemtob,manohar,Praszalowicz:2003ik},
where the quantization condition emerging from the
Wess-Zumino-Witten term selects SU(3) representations of triality
zero~\cite{su3quant}. Concerning the solitonic approach we must
note: i) The soliton approach was used to {\em predict} exotic
states by linking their properties to the known baryons in octet
and decuplet of SU(3)$_{\rm flavor}$. In contrast to the other
treatments it preceded experimental discovery by many years. ii)
Despite some freedom as far as model parameters are concerned the
predictions of the mass were very
accurate~\cite{Praszalowicz:2003ik, Diakonov:1997mm}. While this
accuracy may be viewed as somewhat fortuitous, the low mass of
exotic states in the chiral soliton models is natural
irrespectively of the dynamical details. iii) The width was
predicted to be very small~\cite{Diakonov:1997mm}, which is
consistent with the widths presently observed \cite{Nus,ArAzPo}.
The narrowness of exotic states follows from the cancellation of
different contributions in the decay operator. The width equals
identically zero in the nonrelativistic limit of the model.
Although highly nonintuitive this cancellation provides a natural
(within the model) explanation of the small width of $\Theta^+$
and, simultaneously,  moderately large width of $\Delta^{++}$ for
which no cancellation takes place.

In the present paper we will in the first part review shortly the
solitonic approach to the $\Theta^{+}$ and the antidecuplet. We
will make some comments on the large $N_c$-limit and the problems
of the solitonic (i.e. rigid rotor) approach. Then we discuss
shortly the Jaffe-Wilczek model for pentaquarks and give some
account of the lattice gauge results for $\Theta^{+}$. In the
second part of the paper we apply the \emph{model-independent}
solitonic approach to evaluate the magnetic moments of the
$\Theta^{+}$ and the full antidecuplet. As a byproduct we obtain
also the strange magnetic moment of the nucleon.


\section{Chiral quark-soliton approach}

As well known, the most important feature in QCD from the point of
view of the light hadron structure is the Spontaneous Chiral
Symmetry Breaking (S$\chi$SB): as its result, almost massless
$u,d,s$ quarks get the dynamical momentum-dependent masses
$M_{u,d,s}(p)$, and the pseudoscalar mesons $\pi, K, \eta$ become
light (pseudo) Goldstone bosons. At the same time, the
pseudoscalar mesons are themselves bound $q\bar q$ states. The
interaction of pseudoscalar mesons with constituent quarks is
dictated by chiral symmetry. It can be written in the following
compact form: \beq \mathcal{L}_{\rm eff}=\bar
q\;\left[i\dd-M\exp(i\,\gamma_5\, \pi^A \lambda^A/F_\pi)
\right]\,q,\quad \pi^A=\pi,K,\eta. \la{lagrangian}\eeq

A full solution of this effetive low-energy theory would imply
functional integrals over the quark fields $\psi$ and the
Goldstone fields $\pi^A$. However, in the limit of large $N_c$ it
is known from Witten's work that the baryons are solitons of such
a chiral action. Therefore the soliton as such is not problematic,
the questions is only, whether the large-$N_c$ limit is a good
approximation to the real world with $N_c = 3$, or, at least, if
there are some qualitative features in the large-$N_c$ world which
are reflected in reality.

The model of baryons based on the large $N_c$-limit is the Chiral
Quark-Soliton Model ($\chi$QSM) and variants of it~\cite{CQSM}.
The ``soliton'' is another word for the self-consistent pion mean
field in the nucleon in which the quarks move (in many body
language: Hartree-Fock mean field).
 Since the model operates with
explicit quark degrees of freedom (i.e. it uses as basic object a
regularized many-body Slater determinant) one is able to compute
any type of observables, {\it e.g.} form factors of light baryons,
relativistic quark (and antiquark) distributions, generalized
quark distributions inside nucleons, and the baryon light-cone
wave functions. In contrast to the naive quark models, the
$\chi$QSM is relativistic-invariant. Being such, it necessarily
incorporates quark-antiquark admixtures to the nucleon.
Quark-antiquark pairs appear in the nucleon on top of the $N_c=3$
valence quarks either as particle-hole excitations of the Dirac
sea (read: mesons) or
  as collective excitations
of the mean chiral field, both sorts belonging to the same Hilbert
space. The model has a close similarity to the Skyrme-model.
There, one has integrated out the quark fields and operates with
the action involving directly the $\pi^A$ without the quark
fields. This simplification has a price because  i) in the Skyrme
model one is restricted to local actions, whereas the $\chi$QSM
has the non-local one, ii) the local action is truncated at the 4
derivative order with only one specific term, namely the so-called
Skyrme term retained, in order to stabilize the soliton. One can
formally obtain the effective action of the Skyrme type from the
one of the $\chi$QSM by a gradient expansion which, however,
is poorly converging, if at all, for the soliton configuration. \\

Since the many-body Hartree-Fock states composed of valence- and
sea-quarks are degenerate with respect to rotations in space and
flavor-space these collective degrees of freedom have to be
treated explicitly yielding quantum numbers of the considered
baryon. In the rigid rotor approximation these rotations in space
and flavor-space are described by the collective Hamiltonian
which in case of the SU(3) takes the following form
\cite{Blotz:1992pw}:
\begin{equation}
\hat{H}=\mathcal{M}_{sol}+\frac{J(J+1)}{2I_{1}}+\frac{C_{2}(\text{SU(3)}%
)-J(J+1)-\frac{N_{c}^{2}}{12}}{2I_{2}}+\hat{H}^{\prime},
\label{Eq;1}
\end{equation}
Here $\mathcal{M}_{sol}$ is the classical mean-field energy of the
quark system. Since the strange quarks have a finite mass $m_s$ we
have to
add a symmetry breaking term given by:%
\begin{equation}
\hat{H}^{\prime}=\alpha D_{88}^{(8)}+\beta Y+\frac{\gamma}{\sqrt{3}}%
D_{8i}^{(8)}\hat{J}_{i},\label{Hsplit}%
\end{equation}
where parameters $\alpha$, $\beta$ and $\gamma$ are of order
$\mathcal{O}(m_{s})$. Here $D_{ab}^{(\mathcal{R})}(R)$ denote
SU(3) Wigner rotation matrices and $\hat{J}$ is a collective spin
operator. The Hamiltonian given in Eq.(\ref{Hsplit}) acts on the
space of baryon wave functions $\left|
\mathcal{R}_{J},B,J_{3}\right\rangle $:
\begin{equation}
\left|  \mathcal{R}_{J},B,J_{3}\right\rangle =\psi_{(\mathcal{R}%
;Y,T,T_{3})(\mathcal{R}^{\ast};-Y^{\prime},J,J_{3})}=\sqrt{\mathrm{dim}%
(\mathcal{R})}(-1)^{J_{3}-Y^{\prime}/2}D_{Y,T,T_{3};Y^{\prime},J,-J_{3}%
}^{(\emph{R})\ast}(R).\label{Eq:wave_f}%
\end{equation}
Here, $\mathcal{R}$ stands for the allowed irreducible
representations of the SU(3)$_{\rm flavor}$ group, and
$Y,T,T_{3}$ are the corresponding hypercharge, isospin, and its
third component, respectively. Due to the presence of the
Wess-Zumino term in the imaginary part of the effective action the
right hypercharge $Y^{\prime}=1$ is constrained to be unity for
the physical spin states for which $J$ and $J_{3}$ are spin and
its third component. The Wess-Zumino constraint selects a tower of
allowed SU(3)$_{\rm flavor}$ representations: the lowest ones,
octet and decuplet, coincide with the representation content of
the quark model. This has been considered as a success of the
collective quantization and as a sign of certain duality between
rigidly rotating heavy soliton and constituent quark model. A
third lowest representation is antidecuplet which has been
considered as an artifact of the model and therefore disregarded
until the work of Diakonov, Petrov and Polyakov
\cite{Diakonov:1997mm}.

The procedure of Refs.\cite{Diakonov:1997mm} and
\cite{Praszalowicz:2003ik} consists now in using Eqs.~(\ref{Eq;1})
with $N_c=3$ and (\ref{Hsplit}) and in the present case analogous
equations for the decays and determining the model parameters such
as $I_1,I_2,\alpha ,\beta, \gamma$ and the other parameters from
experimental data.

Taking into account recent experimental observations of the mass
of the $\Theta^{+}$, the parameters entering Eq.(\ref{Hsplit}) can
be conveniently parameterized solely in terms of the pion-nucleon
$\Sigma_{\pi
  N}$ term (assuming $m_{s}/(m_{u}+m_{d})=12.9$) as
\cite{Praszalowicz:2004dn}:
\begin{equation}
\alpha=336.4-12.9\,\Sigma_{\pi N},\quad\beta=-336.4+4.3\,\Sigma_{\pi N}%
,\quad\gamma=-475.94+8.6\,\Sigma_{\pi N}\label{albega}%
\end{equation}
(in units of MeV). Moreover, the inertia parameters which describe
the
representation splittings%
\begin{equation}
\Delta M_{{10}-8}=\frac{3}{2I_{1}},\;\;\Delta M_{{\overline{10}}-8}%
=\frac{3}{2I_{2}}%
\end{equation}
take the following values (in MeV):%
\begin{equation}
\frac{1}{I_{1}}=152.4,\quad\frac{1}{I_{2}}=608.7-2.9\,\Sigma_{\pi
N}.\label{ISigma}%
\end{equation}
Equations (\ref{albega}) and (\ref{ISigma}) follow from the fit to
the masses of the octet and decuplet baryons as well as that of
the $\Theta^{+}$.  If, furthermore, one imposes the additional
constraint that $M_{\Xi_{\overline{10}}}=1860$ MeV, then
$\Sigma_{\pi N}=73$ MeV \cite{Praszalowicz:2004dn} (see also
\cite{Schweitzer:2003fg}) in agreement with recent experimental
estimates~\cite{Sigma}.

Since the symmetry-breaking term of the collective Hamiltonian
(\ref{Hsplit}) mixes different SU(3) representations, the
collective wave functions are given as the following linear
combinations
\cite{Kim:1998gt}%
\begin{align}
\left|  B_{8}\right\rangle  &  =\left|  8_{1/2},B\right\rangle
+c_{\overline {10}}^{B}\left|  \overline{10}_{1/2},B\right\rangle
+c_{27}^{B}\left|
27_{1/2},B\right\rangle ,\nonumber\\
\left|  B_{10}\right\rangle  &  =\left|  10_{3/2},B\right\rangle +a_{27}%
^{B}\left|  27_{3/2},B\right\rangle +a_{35}^{B}\left|
35_{3/2},B\right\rangle
,\nonumber\\
\left|  B_{\overline{10}}\right\rangle  &  =\left|  \overline{10}%
_{1/2},B\right\rangle +d_{8}^{B}\left|  8_{1/2},B\right\rangle +d_{27}%
^{B}\left|  27_{1/2},B\right\rangle +d_{\overline{35}}^{B}\left|
\overline{35}_{1/2},B\right\rangle, \label{admix}
\end{align}
where $\left|  B_{\mathcal{R}}\right\rangle $ denotes the state
which reduces
to the SU(3) representation $\mathcal{R}$ in the formal limit $m_{s}%
\rightarrow0$ and the spin index $J_{3}$ has been suppressed. The
$m_{s}$-dependent (through the linear $m_{s}$ dependence of
$\alpha$, $\beta$ and $\gamma$) coefficients in Eq.(\ref{admix})
are documented in \cite{YangKim}.\\

A formalism similar to the present one has been used by
Prasza{\l}owicz \cite{Praszalowicz:2003ik} in 1987 to make the
prediction for the energy of the $\Theta ^+$, with, however, the
simplified splitting Hamiltonian (\ref{Hsplit}) where
$\beta=\gamma=0$\footnote{Note that $\beta$ and $\gamma$ are $N_c$
suppressed with respect to $\alpha$.}. In fact, since the mass of
$\Theta ^+$ could not be used as an input parameter, he used
corrections quadratic in $m_s$ to fit the non-exotic mass
spectrum. With his original choice of parameters for the
pion-nucleon sigma term $\Sigma_{\pi N}=60$ MeV and the stange
quark mass $m_s=200$ MeV he has predicted the mass of the $\Theta
^+$ to be $1534$ MeV.

Diakonov, Petrov and Polyakov \cite{Diakonov:1997mm} used exactly
the above formalism  with $H^{\prime}$ given by Eq.(\ref{Hsplit}),
however, they assumed the state $N^* (1710)$ to be a member of the
proposed anti-decuplet in order to have a prediction for the
absolute mass of $\Theta ^+$. They calculated in addition the
decay width $\Gamma _{\Theta ^+}<15$ MeV, which triggered the
experimental searches. In this estimate they start from the
experimental value of  $g_{\pi N N}$ and of the singlet axial-vector
current matrix element of the nucleon. In their formalism the
narrowness of $\Theta ^+$ occurs due to a cancellation of the
couplings in the collective decay operator as a conspiracy of the
SU(3)- group-theoretical factors and phenomenological values of
these couplings. The cancellation is, however, by no means
accidental. Indeed, in the small soliton limit (corresponding to
the non-relativistic limit) the cancellation is exact.

Actually Ellis, Karliner and Prasza{\l}owicz
\cite{Praszalowicz:2004dn} reanalyzed the above predictions. The
calculated ranges of the chiral-soliton moment of inertia $I_2$
and the phenomenological ranges of $\Sigma_{\pi N}$ were used
together with the observed baryon octed and decuplet mass
splittings to estimate $1430$ MeV $<$ $m_{\Theta^+}$ $<$ $1660$
MeV and $1790$ MeV $<$ $m_{\Xi^{--}}$ $<$ $1970$ MeV. The authors
emphasize that more precise predictions rely on ambiguous
identifications of non-exotic baryon resonances. If they take the
 singlet axial-vector-current matrix element from deep-inelastic lepton
scattering experiments they find a suppression of the total
$\Theta ^+$ and $\Xi^{--}$ decay widths which, however, may
not be sufficient by itself to reproduce the narrow width required
by experiment. If they calculate in addition SU(3) breaking
effects due to representation mixing they find that those tend to
supress the $\Theta ^+$ decay width while enhancing that of
$\Xi^{--}$\\

In a recent paper Diakonov and Petrov \cite{diakpetmismem} gave a
rather intriguing reason ~\cite{diakpetrwav} for the small width
of $\Theta ^+$
 based on the formalism of
Ref.~\cite{petrpolwav}. The argument is fully based on the fact
that the $\Theta ^+$ decays into kaons, which are Goldstone
bosons, although there will be some corrections due to the finite
mass of the kaon. In the picture of a rotating soliton,
pentaquarks are baryons consisting of valence quarks and quarks in
the polarized Dirac sea. The polarization of the Dirac sea
corresponds to quark-antiquark excitations of the unpolarized
Dirac sea (vacuum). Even if the three `valence' quarks are not too
relativistic, the quark-antiquark pair inside the pion or kaon
mean field is always relativistic. The non-relativistic
wave-function description of pentaquarks makes no sense.
``Measuring" the quark position with an accuracy higher than the
pion Compton wave length of $1\,{\rm fm}$ produces a new pion,
{\it i.e.} a new quark-antiquark pair. What makes sense in this
situation is describing baryons in the infinite momentum frame. In
that frame there can be no production and annihilation of quarks,
and the baryon wave function falls into separate sectors of the
Fock space: three quarks, five quarks, etc. The difference between
the ordinary nucleon and $\Theta^+$ is that the nucleon has a
three-quark component (but necessarily has also a five-quark
component) while $\Theta$'s Fock space starts from the five-quark
component. One can now consider the decay amplitude of $\Theta$ or
$\Xi_{\overline{10}}$ into an octet baryon and a pseudoscalar
meson. Owing to the Goldberger--Treiman relation it is equivalent
to evaluating the matrix element of the axial charge between the
antidecuplet and the octet baryons. In the infinite momentum frame
the axial charge does not create or annihilate quarks but only
measures the transition between the existing quarks. Therefore,
the matrix element in question is non-zero only between the
pentaquark and the {\em five-quark} component of the octet baryon.
Hence it is suppressed to the extent the five-quark component of a
nucleon is less than its three-quark component. There is an
additional suppression owing to the specific flavor structure of
the nucleon's five-quark component where the antiquark is in a
flavor-singlet combination with one of the four quarks
\cite{Diakonov:1997mm}. This is why, qualitatively, the exotic
$\Theta^+$ is narrow. Actually, one should not conclude from these
arguments that necessarily all antidecuplet baryons have a width
as narrow as $\Theta ^+$ or that the baryons of the 27-plet are
even narrower. In fact the number and kind of decay channels play
a role as well as the phase space.


\section{Large $N_c$-arguments}


The solitonic approach to $\Theta ^+$, however successful it is,
is based on an ansatz and as such one must check the
self-consistency of the approach. Ultimately, following arguments
by Witten, it is justified at large $N_c$ by two reasons: First,
looking e.g. at Eq.(\ref{lagrangian}) a calculation of baryonic
properties needs functional integrals over both bosonic ($\pi^A$)
and fermionic ($\psi$) fields. In order to reduce the problem the
stationary phase approximation is invoked which results in
omitting the bosonic functional integral and using instead just
the one bosonic field which minimizes the effective action of the
system. This is only justified in large $N_c$, where fluctuations
of the bosonic field are suppressed. Second: In quantizing the
rotations in space and flavor space one assumes the rotational
motion to be adiabatic. This means that angular velocities go like
$J/{ I} \sim N_c^{-1}$ yielding frequencies (and hence excitation
energies) of order $N_c^{-1}$.  This, in turn, implies a
Born-Oppenheimer separation of the slow collective rotational
motion from the faster modes associated with vibrations of the
meson fields (with time scales of order $N_c^0$ and hence energies
of order $N_c^0$). Because of this scale separation the collective
rotational modes can be quantized separately from the intrinsic
vibrations. Actually this procedure has been applied with great
success to many properties of the baryon octet and decuplet and
has been used for the predictions of $\Theta ^+$ and the
anti-decuplet.

In several papers Cohen \cite{cohenlargenc} has criticized the
application of the solitonic approach with rigid-rotor quantization
to $\Theta^+$, which is based apparently on SU(3) solitons.
Extending the collective rotational quanization from SU(2) to
SU(3)$_{\rm flavor}$ is straightforward. The only essential new
feature is the inclusion of the topological Wess-Zumino-Witten
term which builds in the anomalies. For simplicity, first consider
the case of exact SU(3) symmetry. The standard semiclassical
approach is then to first solve the problem classically which
yields a hedgehog configuration in a two-flavor subspace (which
we will take to an intrinsic u-d subspace by convention).
Following Witten's ansatz one introduces a time-dependent,
space-independent, global SU(3) rotation $A(t)$. At this stage the
parallel of the two flavor case is virtually exact.  However, the
Wess-Zumino-Witten term introduces a constraint: In the body-fixed
(co-rotating) frame the hypercharge must be $N_c/3$.  This
constraint plays a critical role in what follows.

Going through this procedure yields the collective rotation
Hamiltonian of the form given in Eq.~(\ref{Eq;1}) where $I_1$
($I_2$) is the moment of inertia in the body fixed frame for
motion within (out of) the original SU(2) subspace. This procedure
yields masses given by
\begin{equation}
\label{mass}
M=M_0+\frac{J(J+1)}{2I_{1}}+\frac{C_{2}(\text{SU(3)}%
)-J(J+1)-\frac{N_{c}^{2}}{12}}{2I_{2}},
\end{equation}
with
\begin{equation}
C_2 =  \left( p^2 + q^2 + p q + 3(p +q)\right )/3  \; .
\end{equation}
 $C_2$ is  the Casimir operator, and is labeled  by $p,q$
which specify the SU(3) representation.  As mentioned in the
previous section the constraint due to the Wess-Zumino-Witten term
imposes the restriction that the representation must have a state
with $Y=N_c/3$ and implies that angular momentum is determined by
the condition that $2J+1$ equals the number of states with strangeness S=0.
Plugging in $N_c=3$, one sees the lowest representations are
$(p,q)=(1,1)$ (spin 1/2 octet), $(p,q)=(3,0)$ (spin 3/2 decuplet)
and $(p,q)=(0,3)$ (spin 1/2 anti-decuplet).  This last
representation is clearly exotic.

The question addressed by Cohen \cite{cohenlargenc} and Pobylitsa
\cite{pobylitsalargenc} is whether the rigid-rotor type
semiclassical projection used here is correct if applied to exotic
states. Since the question ultimately boils down to whether the
Born-Oppenheimer separation is justified at large $N_c$ for these
states, care should be taken to keep $N_c$ arbitrary and large
throughout the analysis.  In particular, at least when doing
formal studies of the $N_c$ dependence, one ought not set $N_c=3$
when imposing the constraint of the Witten-Wess-Zumino term as
done by basically all soliton practitioners, who assume at the
very beginning $Y^{\prime}=1$. In doing this one sees that the
flavor SU(3) representations at large $N_c$ differ from their
$N_c=3$ counterparts. The lowest-lying representation consistent
with the Witten-Wess-Zumino term constraint has $ \left (p,q
\right ) = \left( 1, \frac{N_c-1}{2} \right) $ and has $J=1/2$.
Thus, it is a clear analog of the octet representation and will be
denoted as the ``8'' representation. (The quotes are to remind us
that it is {\it not} in fact an octet representation.)  Similarly
the next lowest representation,  has $ \left ( p,q \right ) =
\left( 3,\frac{N_c-3}{2} \right ) $ and has $J=3/2$.  It is the
large $N_c$ analog of the decuplet and will be denoted ``10''. The
salient feature of the anti-decuplet is that its lowest
representation contains an exotic S=+1 state.  Thus its large
$N_c$ analog is the lowest representation containing manifestly
exotic states.  This representation is $ \left ( p,q \right ) =
\left( 0, \frac{N_c+3}{2} \right ) $ and has $J=1/2$; it will be
denoted ``$\overline{10}$''.

It is now interesting to consider the excitation energy of the
exotic states which can be computed from Eq.~(\ref{mass}):
\begin{equation}
 M_{``\overline{10}"} - M_{``8"}  = \frac{3 + N_c}{4 I_2} \; .
 \label{10bar8quote}
\end{equation}
Noting that $I_2 \sim N_c$, one sees that this implies that the
excitation energy of the exotic state is of order $N_c^0$.  This
may be contrasted to the excitation of the non-exotic ``10''
representation which is of order $1/N_c$, since the spin $J\sim
N_c^0$:
\begin{equation}
 M_{``{10}"} - M_{``8"}  = \frac{3}{2 I_1} \; .
 \label{108quote}
\end{equation}

The fact that the standard semiclassical quantization gave excitation
energies of order $N_c^0$ for exotic states means that the
approach is not justified for such states unless further arguments
can be invoked . Recall that the analysis was justified
self-consistently via a Born-Oppenheimer scale separation.  For
the non-exotic states this is justified. However, for the exotic
states the characteristic time associated with the excitations
(one over the energy difference) is of order $N_c^0$. This is
the same characteristic time as the ``fast'' vibrational
excitation of the meson fields.  One cannot therefore justify
treating the rotational ``collective'' degrees of freedom
separately. Thus, following the criticisms of Cohen
\cite{cohenlargenc} and Pobylitsa \cite{pobylitsalargenc}, the
prediction of $\Theta^+$ properties via this collective
quantization procedure cannot be justified from large $N_c$.

 Actually Diakonov and Petrov \cite{diakpetlargenc} have in
fact constructed the generalization of octet, decuplet,
antidecuplet, 27-plet, 35-plet and further multiplets to the case
of arbitrary $N_c$. They classify these multiplets by a quantum
number called ``exoticness", i.e. the number of extra
quark-antiquark pairs needed to compose the multiplet. The
splittings between masses of the multiplet with the same
exoticness are $O(1/N_c)$ {\it i.e.} parametrically small at large
$N_c$. The spectrum of multiplets with growing exoticness is
equidistant with an $O(1)$ spacing. On the one hand this spacing
can be interpreted as an energy for adding a quark-antiquark pair
but on the other hand it is a rotational excitation. Despite that
it becomes comparable to vibrational or radial excitations of
baryons, both non-exotic and exotic bands are, at large $N_c$,
described as collective rotational excitations of the ground-state
baryons: the corrections due to coupling between rotations and
vibrations die out as $1/N_c$. The collective rotational
quantization description fails only when the exoticness becomes
comparable to $N_c$. The newly discovered $\Theta^+$ baryon
belongs to the exoticness=1 in the antidecuplet. The larger $N_c$
the more accurate would be its description as a rotational state
of a chiral soliton. Diakonov and Petrov \cite{diakpetlargenc}
support their estimates by considering a simple model consisting
of a charged particle in the field of a monopole. They take this
example from arguments of Witten and of Guadagnini, who stress the
apparent formal similarity of the physics of this model to the
Wess-Zumino term. In fact all relationships of Diakonov and Petrov
\cite{diakpetlargenc} derived in the general case are born out in
this model. In particular the coupling between vibrations and
rotations vanish in the limit of what is in the model the analogue
of large $N_c$. However, if one generalizes the model by
considering two charged particles interacting by a harmonic
potential and moving in the field of a monopole the coupling of
rotational and vibrational degrees of freedom of this system is by
no means vanishing but strong \cite{pobylitsalargenc}.

Large-$N_c$ arguments apply also to the width of the baryons
considered, because if the approach is justified, then at a formal
level the width must approach zero at large $N_c$.  Of course, for
non-exotic states such as the decuplet, this is true. The reason
is simply phase space. As $N_c$ grows, the excitation energy for
these states drops thereby reducing the phase space for decay to
zero. In contrast, as shown recently by Prasza{\l}owicz \cite{Pra}
the width of the $\Theta^+$ as calculated via the standard
collective rotational approach is of the order $N_c^0$. This
demonstrates that the procedure is not self consistent on the
basis of pure large-$N_c$ arguments.

Actually the arguments concerning the decay width of $\Theta^+$
are extremely subtle and by no means fully understood, as the
following consideration demonstrates: The decay width of a state
$B$ in unbroken SU(3)$_{\rm flavor}$ representation $R$ into an
octet baryon $B^{\prime }$ and a pseudoscalar octet meson $\varphi
$ reads:
\begin{equation}
\Gamma _{B\rightarrow B^{\prime }\varphi }=\frac{G_{R}^{2}}{8\pi
M\,M^{\prime }}\times C^{R}(B,B^{\prime },\varphi )\times
p_{\varphi }^{3}. \label{dw}
\end{equation}%
Here $M$, $M^{\prime }$ and $p_{\varphi }$ are baryon masses and
meson momentum in the decaying baryon reference frame
respectively. $C^{R}(B,B^{\prime },\varphi )$ is the relevant
SU(3) Clebsch-Gordan coefficient (squared) and $G_{R}$ the decay
constant. In order to calculate $\Gamma$ we take for baryon masses
and phase-space factor $p_{\varphi}$ {\em physical} values, rather
than model predictions obtained in a given order of $1/N_c$ and
$m_s$. In that way we effectively sum up the infinite  series in
these small parameters leaving the secure island of theoretical
paradise. The Clebsch-Gordan factor $C^{R}(B,B^{\prime },\varphi )$
appears in any model with SU(3)$_{\rm flavor}$ symmetry and is a
rational function in $N_c$ because of the representation choice
discussed above. The genuine model dependence is buried in the
reduced matrix elements $G_R$.

Now: $G^2_{10,\overline{10}}$ scale as $N_c^3$ (unless the exact
cancellation for antidecuplet takes place and
$G_{\overline{10}}\equiv 0$). It has been shown that
$C^{10}(\Delta,N,\pi )\sim N_c^0$, while
$C^{\overline{10}}(\Theta,N,K )\sim 1/N_c$. So the amplitude
(squared) for the decay of $\Theta$ is suppressed with respect to
that of $\Delta$. However, momenta of outgoing mesons scale
differently with $N_c$ in the chiral limit because of
Eqs.(\ref{10bar8quote},\ref{108quote}):
\begin{eqnarray}
p_{\pi}\sim 1/N_c & ~ & p_K \sim N_c^0. \label{Ncscale}
\end{eqnarray}
However, in Nature neither $\pi$ nor K mesons are massless and
both $p_{\pi}$ and $p_K$ are of the same order of 230 MeV and the
scaling (\ref{Ncscale}) does not hold. So for $m_{\pi} \ne 0$ and
$m_K \ne 0$ both meson momenta scale as $N_c^0$, however $\Delta$
does not decay, because in the large $N_c$ limit it is degenerate with
a nucleon and cannot emit a massive particle, whereas $\Theta^+$
does decay. It is an instructive example how subtle is an
interplay of theoretical limits $N_c \rightarrow \infty$ and $m_q
\rightarrow 0$.\\

In summary, the coupling between rotational and vibrational
degrees of freedom seems to be noticeable in general and no
systematic large-$N_c$ arguments can be found that it should be
small. This is consistent with results of the Skyrme model on
rotation-vibration mixing \cite{Weigel}. This is also consistent
with arguments based on non-relativistic many-body theory, where
one-particle--one-hole excitations and the vibrational states of
the random-phase approximations in fact span the same
Hilbert-space. Thus, the question arises why one does not drop the
solitonic apprach with rigid-rotor quantization and uses only pure
large-$N_c$ arguments. In such a case the energy can be described
but the width is a problem. Large-$N_c$ estimates yield a width
$\Gamma \sim O(1)$ and hence of the same order as the excitation
energy. The experiments, however, measure a width $\Gamma < 10$
MeV which is very much smaller than the excitation energy and does
not fit into any known $N_c$ ordering. Thus, if the width of the
$\Theta ^+$ is really small, as the experiments indicate, there
must be particular dynamic reasons for the smallness, which exist
on top of what is required for the validity of the large-$N_c$
expansion. What this dynamics actually looks like and what the
underlying structures of the quark and gluon fields are and how
far this is reproduced by the rigid-rotor quantization is
certainly the subject of future investigations.

It is worthwhile to mention that in a recent sum rule calculation
~\cite{joffeogan} it was observed that the decay width of
$\Theta^+$ is strongly suppressed in the chiral limit. This result
indicates that an understanding of the narrow width of $\Theta^+$
does require to take carefully into account both the chiral limit
and the large-$N_c$ limit. This is a subtle issue, however,
bearing in mind that the two limits do not comute.

For completeness one should note also another perspective on
problems with the standard method used to quantize these solitons,
which is given by Klebanov et al.~\cite{klebanov}. These authors
reconsider the relationship between the $SU(3)$ rigid-rotator and
the bound-state approaches to strangeness in the chiral soliton
models. For non-exotic ${\rm S}=-1$ baryons the bound-state approach
matches for small kaon mass $m_K$ onto the rigid-rotator approach,
and the bound-state mode turns into the rotator zero-mode.
However, for small $m_K$, there are no ${\rm S}=+1$ kaon bound states or
resonances in the spectrum. This shows that for large $N_c$ and
small $m_K$ the exotic state is an artifact of the rigid-rotator
approach. An ${\rm S}=+1$ near-threshold state with the quantum numbers
of the $\Theta^+$ pentaquark comes into existence only when
sufficiently strong SU(3) breaking is introduced into the chiral
Lagrangian. Therefore the authors argue that pentaquarks are not
generic predictions of the chiral soliton models.

\section{Quark models}
\label{quarkmodels}

Perhaps the most popular class of models for the structure of
$\Theta ^+$ and its relatives consists of variants of the
constituent quark model \cite{JW,qm,Shuryak:2003zi}. Here the most
prominent one is the model introduced by Jaffe and
Wilczek~\cite{JW} although these authors calculate only energies
and no decay widths. Jaffe and Wilczek describe the exotic states
as pentaquarks composed of two strongly bound scalar diquarks
$(ud)$ interacting with a strange anti-quark. The model predicts
that the lowest-lying pentaquark states are a nearly ideally mixed
combination of an SU(3)$_{\rm flavor}$ octet and an SU(3)$_{\rm
flavor}$ anti-decuplet (both consisting of pentaquarks). One basic
feature of the model is the prediction of two pentaquark states
with nucleon quantum numbers: one with a mass of $\sim 1450$ MeV
and the other with a mass of $\sim 1700$ MeV. These states have
been tentatively identified with the Roper resonance ${\rm
N}^*(1440)$ and the ${\rm N}^*(1710)$. Actually, there exists an
investigation by Cohen \cite{CohenJW} which focuses on the widths
of the $N^*$ states with emphasis on determining whether the
Jaffe-Wilczek model is consistent with presently known
phenomenological constraints. Cohen has shown that the
identification of the two nucleon states as the Roper resonance
and the ${\rm N}^*(1710)$ is inconsistent with the phenomenology
of the widths given the assumptions underlying the model and
recent bounds on the $\Theta^+$ width ~\cite{Nus,ArAzPo}. The
reason for this inconsistency in the identification is easy to
see. Given the assumption of nearly ideal mixing, the two physical
states with nucleon quantum numbers each have substantial
contributions from both the octet and anti-decuplet  multiplets.
The  $\Theta^+$, which has no octet contribution, is extremely
narrow as all experiments show, which found this particle. Hence
the coupling of the anti-decuplet part of the two states to the
pion plus nucleon final state is quite small. In contrast, the
Roper resonance is known to be very broad with a large coupling to
the pion plus nucleon state.  If the Roper resonance is identified
with the lighter of the two ${\rm N}^*$ states in the ideal mixed
octet plus anti-decuplet, one can deduce that the coupling to the
octet component must be large.  A large coupling to the octet, in
turn, implies a large partial width for the higher of the two
${\rm N}^*$ states. However, the partial width for the ${\rm
N}^*(1710)$ into pion plus nucleon is, in fact, quite small. It is
interesting to note that Cohen's arguments are purely based on the
phenomenological predictions of the Jaffe-Wilczek model without
making any judgement of the underlying picture of strongly bound
scalar diquarks interacting with an anti-quark in a particuar
state with $L=1$. (For a discussion of alternative diquark-models
providing also an estimate of the width of $\Theta ^+$ see e.g.
Refs.~\cite{Shuryak:2003zi} and~\cite {Hong:2004xn})

The simple arguments above are detailed as follows: The only basic
assumption about the dynamics which goes into the derivation of
Cohen is that the decay of an unstable particle can be cleanly
treated via the coupling of the unstable particle to the continuum
of open channels with no significant interference from background
effects. In this circumstance, the partial width for the two-body
decay of a $(1/2)^+$ baryon ($B'$) into another baryon ($B$) plus
a pseudo-scalar meson ($\varphi$) may be computed by combining a
coupling constant parameterizing the strength with appropriate
kinematic factors which incorporates both the phase space and the
p-wave nature of the coupling~\cite{Shuryak:2003zi} of a
pseudo-scalar meson to $(1/2)^+$ states (\ref{dw}).

Here we shall relate the decays of three baryons, the $\Theta^+$
(into a kaon plus a nucleon) and the two ideally mixed pentaquarks
with nucleon quantum numbers (into a pion plus a nucleon). The
ideally mixed  states will be labeled $N_0^{\ast}$ and
$N_1^{\ast}$ where the subscript indicates the number of strange
quarks (which equals the number of anti-strange quarks in the
state) so that with the tentative identification of Jaffe and
Wilczek, the $N_0^{\ast}$ is the Roper resonance and the ${\rm
N}_1^{\ast}$ is the ${\rm N}^*(1710)$:

\begin{align}
\mathrm{Roper}& \rightarrow \left| N_{0}^{\ast }\right\rangle =\sqrt{\frac{1%
}{3}}\left| \overline{10},N\right\rangle -\sqrt{\frac{2}{3}}\left|
8,N\right\rangle ,  \nonumber \\
{\rm N}^{\ast }(1710)& \rightarrow \left| N_{1}^{\ast }\right\rangle =\sqrt{\frac{2%
}{3}}\left| \overline{10},N\right\rangle +\sqrt{\frac{1}{3}}\left|
8,N\right\rangle ,  \label{Rop1710}
\end{align}%
The amplitude for the decay width  of %
$N_{0,1}^{\ast}$ would be the sum over $R=\overline{10}$ and 8.
However, since the decay width of $\Theta ^{+}=\left|
\overline{10},\Theta ^{+}\right\rangle $
is very small, indicating that $G_{\overline{10}%
}\sim 0$ (remember $\Theta^+$ does not mix), the $\overline{10}$
component in Eqs.(\ref{Rop1710}) can be safely neglected in the
first approximation. Then
\begin{equation}
\frac{\Gamma _{N_{0}^{\ast }\rightarrow N\pi }}{\Gamma
_{N_{1}^{\ast }\rightarrow N\pi }}\sim
2\frac{M_{1}}{M_{0}}\frac{p_{0}^{3}}{p_{1}^{3}}\sim 0.75
\label{Gamrat}
\end{equation}%
where we have used physical masses for Roper and ${\rm N}^{\ast
}(1710)$. Equation (\ref{Gamrat}) indicates that partial decay
widths of the nucleon-like pentaquark resonances have to be of the
same order: either both small or both large. This observation was
first made by Cohen \cite{CohenJW} and presented in form of the
inequality connecting different decay constants. Experimentally
\cite{CohenJW} $\Gamma _{\text{Roper}\rightarrow N\pi }\sim 228$
MeV and $\Gamma _{1710\rightarrow N\pi }\sim 15$ MeV. Order of
magnitude difference between the partial decay widths of Roper and
${\rm N}^{\ast }(1710)$ remains in contradiction with the ideal
mixing scenario.

If one accepts the validity of the model and the reported bounds
on the $\Theta^+$ width ~\cite{Nus,ArAzPo} and goes tentatively
beyond the presently known features of the excited nucleon states
then, following Cohen \cite{CohenJW} one of the following three
scenarios must be adopted: i) the ${\rm N}^*(1710)$ state is, in
fact, a nearly ideally mixed pentaquark and it has a previously
undiscovered narrow partner with a mass in the neighborhood of
$\rm 1450$ MeV; ii) the Roper resonance is a nearly ideally mixed
pentaquark and it has a previously undiscovered broad partner with
a mass in the neighborhood of $\rm 1700$ MeV;  iii) neither the
lower nor the upper ${\rm N}^*$ states have been
detected.\\

It is interesting to note that results, which are in certain
aspects similar to those of Jaffe and Wilczek, are obtained by
dynamical studies in the constituent quark model of Stancu
\cite{stancumix}. There the short-range interaction has a
flavor-spin structure yielding, besides a $\Theta^+$ with positive
parity, a nearly ideal mixing. The masses of the excited nucleon
states obtained are 1451 MeV and 1801 MeV, respectively. Here the
mixing angle is $\theta_N = 35.34^0$, which means that the
``mainly antidecuplet'' state $N_5$ is  67 \%
 $N_{\overline {10}}$
and 33 \% $N_{8}$, and the ``mainly octet'' $N^*$ the other way
round. The latter, having a structure of $q^4 \bar q$, is located
in the Roper resonance mass region at 1430 - 1470 MeV.\\

It remains to be seen if present or future experiments support
these predictions or not. See e.g. some new results from GRAAL
\cite{Kuznetsov}.


\section{Lattice MONTE CARLO calculations for $\Theta ^+$}

It might seem surprising that of the more than 200  theoretical
papers devoted to the subject of exotic baryons in the past year,
there were only four lattice papers\footnote{During preparation of
this paper four other lattice papers appeared~\cite{newlattice}; for
summary see the review by S.Sasaki \cite{Sasaki:2004vz}.}. This
apparent paradox is in detail discussed in a recent report by
Csikor, Fodor, Katz and Kovacs \cite{Fodor}. According to these
authors it has its reason probably in the difficulties and
pitfalls of the lattice approach applied to $\Theta ^+$. Probably
the biggest challenge lattice pentaquark calculations face is how
to choose the baryonic operators. Not only the errors, but also
the very possibility to identify certain states depends crucially
on the choice of operators. Unfortunately there is very little
guidance here and many technical restrictions. Furthermore, since
the five-quark bound states we want to study can be close to the $\bar{\rm K}{\rm N}$
threshold, it is essential in any lattice spectroscopy calculation
to reliably distinguish between genuine five-quark bound states
and meson-baryon scattering states. Both points are discussed in
the present section following the arguments of Ref.~\cite{Fodor}.

In hadron spectroscopy one would like to identify hadronic states
with given quantum numbers. Practically this means the following.
We compute the vacuum expectation value of the Euclidean
correlation function $\langle 0| \CO(t) \CO^\dagger(0) |0 \rangle$
of some composite hadronic operator $\CO$. The operator $\CO$ is
built out of quark creation and annihilation operators. In
physical terms the correlator is the amplitude of the ``process''
of creating a complicated hadronic state described by $\CO$ at
time $0$ and destroying it at time $t$.

After inserting a complete set of eigenstates $|i \rangle$ of the
full QCD Hamiltonian the correlation function can be written as
\begin{equation}
  \langle 0| \CO(t) \CO^\dagger(0) |0 \rangle =
  \sum_i \;\; |\; \langle i | \CO^\dagger(0) |0 \rangle \; |^2 \;
   \; \mbox{e}^{-(E_i-E_0)t},
     \label{eq:corr}
\end{equation}
where
\begin{equation}
  \CO(t) = \mbox{e}^{-Ht}\; \CO(0) \; \mbox{e}^{Ht}
\end{equation}
and $E_i$ are the energy eigenvalues of the Hamiltonian.

Note that since we work in Euclidean space-time (the real time
coordinate $t$ is replaced with $-it$), the correlators do not
oscillate, they rather die out exponentially in imaginary time. In
particular, after long enough time only the lowest (few) state(s)
created by $\CO$ give contribution to the correlator. The energy
eigenvalues corresponding to those states can be extracted from
exponential fits to the large $t$ behaviour of the correlator.

In the simplest case one is typically interested in hadron
masses. A trivial but most important requirement in the choice of
$\CO$ is that it should have the quantum numbers of the state we
intend to study. Otherwise the overlap $\langle i | \CO^\dagger(0)
|0 \rangle$ is zero and the corresponding exponent cannot be
extracted. In order to have optimal overlap with only one state
$|i\rangle$, $\CO^\dagger(0) |0 \rangle$ should be as ``close'' to
$|i\rangle$ as possible.

 One of the most important experimentally still unknown quantum numbers
of pentaquark states is their parity. The simplest baryonic
operators do not create parity eigenstates, rather they couple to
both parity channels. Projection to the $+/-$ parity eigenstates
can be performed as
\begin{equation}
 \CO_{\pm} = \frac{1}{2}(\CO \pm  P \CO P^{-1}).
\end{equation}
If the parity of a state is not known, it can be determined by
computing the correlator in both parity channels and deciding
which channel produces a mass closer to the experimentally
observed one.

After all quantum numbers are fixed, there is still considerable
freedom in the choice of $\CO$. This freedom has to be exploited
to ensure maximal overlap of $\CO^\dagger(0) |0 \rangle$ with the
desired state and minimal overlap with close-by competing, but
unwanted states. This is essential because with the wrong choice
of $\CO$ the desired state might be practically undetectably lost
in the noise.  Unfortunately, beyond the quantum numbers there is
usually little if any guidance in the choice of $\CO$ and herein
lies the biggest challenge of lattice pentaquark spectroscopy. It
is almost impossible to {\em disprove} the existence of a given
state. If one cannot detect it with a given operator $\CO$ , it
might just mean that $\CO$ has too small overlap with the desired
state and the signal is lost in the noise. Indeed, even in the
case of the nucleon simple operators are known to have the
correct quantum numbers, but too little overlap with the nucleon
ground state and no nucleon signal can be extracted from their
correlator ~\cite{Sasakiblumohta}. In the case of the pentaquark we
face even an additional difficulty since due to numerical
complications only two different quark sources are used, one for
the light quarks and the other for the strange quark. All four lattice
pentaquark studies have used this simple choice.

Besides the spatial structure of $\CO$ the single quark spin,
color and flavor indices also have to be arranged properly for
$\CO$ to have the desired quantum numbers. Even then the
arrangement of indices is also not unique. An additional
difficulty one faces here compared to conventional three quark
hadron spectroscopy is that index summation becomes exponentially
more expensive if we increase the number of quarks. While with
three quarks this part of the calculation is usually negligible,
even for the simplest five quark operator it takes up around 50\%
of the CPU time. This circumstance restricted the choice of
pentaquark operators so far to the simplest one.

Pentaquark spectroscopy is further complicated by the presence of
two-particle scattering states lying close to the pentaquark
state. Lattice calculations are always performed in a finite
spatial volume, therefore these scattering states do not form a
continuum. They occur at discrete energy values dictated by the
discrete momenta $p_k = 2k\pi/L, k=0,1,...$, allowed in a box of
linear size $L$. In lattice pentaquark computations it is
absolutely essential to be able to distinguish between these
two-particle nucleon-meson scattering states and genuine five
quark bound states.

In fact, the first experimentally found exotic baryon state, the
$\Theta^+(1540)$ lies just about 100~MeV {\it above} the
$\bar{\rm K}{\rm N}$-threshold. This implies that for large enough time
separation the correlation function is bound to be dominated by
the $\bar{\rm K}{\rm N}$-state. However, the mass difference between the
two states is quite small and the mass of the $\Theta^+$ might
still be reliably extracted in an intermediate time window,
provided that
\begin{equation}
 |\langle \Theta^+ | \CO | 0 \rangle| \; \gg \;
        | \langle N+K | \CO | 0 \rangle |.
\end{equation}
Even then, identifying the $\Theta^+$ is still a non-trivial
matter since the $\Theta^+$ ground state is embedded in an
infinite tower of $\bar{\rm K}{\rm N}$ scattering states with relative
momenta allowed by the finite spatial box. Since the parity of the
$\Theta^+$ is unknown, we have to consider both parity channels.
The situation is qualitatively different in the two channels as
reviewed in detail in Ref.~\cite{Fodor}. In fact tremendous care
is required particularly in the negative-parity channel, which is
the preferred one for three of the four lattice pentaquark
calculations.

Finally, for a convincing confirmation of the pentaquark state in
either positive or negative parity channel, one also has to identify the competing
scattering states observing the volume dependence dictated by the
allowed smallest momentum. This would clearly require a finite
volume analysis combined with a reliable method to extract several
low-lying states from the spectrum. Apart from the volume
dependence of the masses, another powerful tool to distinguish
between two-particle and one-particle states is to check the
volume dependence of their spectral weights.

A popular way of identifying more than one low-lying state from
correlators can be characterized in the following way. If there is
a time interval where more than one state has an appreciable
contribution to the correlator, a sum of exponentials can be
fitted as
\begin{equation}
 \langle 0| \CO(t) \CO^\dagger(0) |0 \rangle =
 C_1\mbox{e}^{-E_0t}+C_2\mbox{e}^{-E_1t}+...
\end{equation}
For this method to yield reliable energy estimates for higher
states, one usually needs extremely good quality data. This is
difficult because lattice spectroscopy of hadrons built out of
light quarks involves two extrapolations. Firstly, simulations at
the physical $u/d$ quark masses would presently be prohibitively
expensive, therefore one has to do several calculations with
heavier quarks and then extrapolate to the physical quark masses.
The lightest quarks used in presently available pentaquark studies
correspond to pion masses in the range 180-650~MeV (see
Table~\ref{table}).

\begin{table}[h]
  \centering
\caption{Lattice spacing and smallest pion mass of lattice
pentaquark calculations.}
{ 
\begin{tabular}{cccc} \hline
                       &  action & $a$ (fm)   &  smallest $m_\pi$ (MeV) \\
                 \hline \hline
 Csikor et al. &  Wilson & 0.17-0.09  &  420 \\ \hline
 Sasaki        &  Wilson & 0.07       &  650 \\ \hline
 Liu et al.    &  chiral & 0.20       &  180 \\ \hline
 Chiu \& Hsieh   &  chiral & 0.09       &  400 \\ \hline
\end{tabular} \label{table} }
\vspace*{-13pt}
\end{table}

Secondly, the space-time lattice is not a physical entity, it is
just a regulator that has to be eventually removed to recover
continuous space-time. This implies that physical quantities have
to be computed on lattices of different mesh sizes and
extrapolated to the zero lattice spacing (continuum) limit.
Lattice simulations can differ from one another in many technical
details and it is only the continuum limit of physical quantities
that is meaningful to compare among different simulations.

Having set the stage we can now present the lattice results along
with our interpretation. Four independent lattice pentaquark
studies have been presented. Their main results can be summarized
as follows, following Ref.~\cite{Fodor}.
\begin{itemize}
\item {\em Sasaki}~\cite{Sasaki:2003gi} using double exponential fits, found a state
consistent with the $\Theta^+$ also in the $I^P=0^-$ channel. He
also managed to identify the charmed analogue of the $\Theta^+$
640~MeV above the $DN$ threshold. (The experimentally found
anticharmed pentaquark lies only about 300~MeV above the
threshold.)

\item {\em Csikor, Fodor, Katz and Kovacs}~\cite{Csikor:2003ng}
using a different operator identified a state in the
 $I^P=0^-$ channel with a mass consistent with the experimental $\Theta^+$
 and the lowest mass found in the opposite parity $I^P=0^+$
 channel was significantly higher. Using $2\times 2$ cross correlators an
 attempt was also made to separate the $\Theta^+$ and the lowest nucleon
 kaon state.

\item {\em Liu et al.}\cite{Liu} reported that they
were not able to see any $\bar{\rm K}{\rm N}$-state compatible with the $\Theta^+$ in either
parity isosinglet channel. Although their smallest pion mass was
the closest to the physical one and they use an improved, chiral
Dirac operator, utilized the nucleon$\times$kaon operator and
their lattice is the coarsest of the four studies. On the other
hand they made use of sophisticated
multi-exponential fits with Bayesian priors.%

\item Finally {\em Chiu~\&~Hsieh}~\cite{Chiu:2004gg}, in
disagreement with the first two studies, saw a positive parity
isosinglet state compatible to the $\Theta^+$, whereas the lowest
state they found in the negative parity state was much higher. In
a subsequent paper \cite{Chiu:2004uh} they also identified states
claimed to be charmed counterparts of the $\Theta^+$.
\end{itemize}

The tentative interpretation of this somewhat controversial
situation by Csikor, Fodor, Katz and Kovacs \cite{Fodor} is as
follows: Liu et al.\ used only one operator with exactly the same
index structure as that of the nucleon kaon system. This might
explain why they see only the expected scattering states. The
three remaining studies could be interpreted to have found genuine
pentaquark states. All three agree that the lowest masses in the
two parity channels differ by about 50\%, but they do not agree on
the parity of the $\Theta^+$ state. While Sasaki and Csikor et
al.\ suggest negative parity, Chiu~\&~Hsieh claim positive parity.
According to the interpretation of Chiu~\&~Hsieh they found
different parity because they used a quark action with better
behavior at small quark masses, albeit the same operator as
Sasaki. The pion masses they use $(\geq 400$~MeV) overlaps with
those of Sasaki $(\geq 650$~MeV). In this region, as long as the
same hadron operator is being used, all other hadron masses in the
literature obtained with these two quark actions agree (see
e.g.\cite{Chiu:2004gg}). Thus it is extremely unlikely that the
same operator with different lattice actions produces such vastly
different masses. Following Ref.~\cite{Fodor} a more likely
resolution of this contradiction is that someone might have simply
misidentified the parity. On the one hand, the results of
Chiu~\&~Hsieh and on the other hand, those of Sasaki (and Csikor
et al.) would become compatible with each other if parities were
flipped in one of them. A possible hint for a parity mismatch is
provided by Chiu~\&~Hsieh in their second paper
\cite{Chiu:2004uh}. They considered two operators with opposite
internal parities, but otherwise having exactly the same quantum
numbers. Contrary to physical expectations, their ordering of the
lowest mass states in the two parity channels turned out to depend
on the internal parity of the operator. This suggests, as argued
in \cite{Fodor}, that internal parity might not have been properly
taken into account. Finally we would like to note that at this
stage one can merely offer these speculations and the issue has to
be resolved by independent studies.

\section{Magnetic moments of anti-decuplet in the chiral quark-soliton model}
\label{magmoms}

In this section we report on a recent study ~\cite{YangKim} in the
context of the chiral quark-soliton model. In fact we derive in a
model-independent way the magnetic moments of the pentaquark
states (i.e. baryon anti-decuplet). This study, which is
strictly in the rigid-rotator scheme with linear $m_s$ corrections
will perhaps provide a way to distinguish some of the above
discussed models. It is also helpful for a systematic study of
photo-production or photo-decays of non-exotic pentaquark states
of the anti-decuplet.

In the formalism of the chiral quark-soliton model and the rigid-rotor
quantization including linear $m_s$-terms the collective
operator for the magnetic moments can be parameterized by six
constants. By definition in the {\em model-independent approach}
they are treated as free
~\cite{Kim:1997ip,Kim:1998gt}:%
\begin{align}
\hat{\mu}^{(0)} &  =w_{1}D_{Q3}^{(8)}\;+\;w_{2}d_{pq3}D_{Qp}^{(8)}%
\cdot\hat{J}_{q}\;+\;\frac{w_{3}}{\sqrt{3}}D_{Q8}^{(8)}\hat{J}_{3},\\
\hat{\mu}^{(1)} &  =\frac{w_{4}}{\sqrt{3}}d_{pq3}D_{Qp}^{(8)}D_{8q}%
^{(8)}+w_{5}\left(  D_{Q3}^{(8)}D_{88}^{(8)}+D_{Q8}^{(8)}D_{83}^{(8)}\right)
\;+\;w_{6}\left(  D_{Q3}^{(8)}D_{88}^{(8)}-D_{Q8}^{(8)}D_{83}^{(8)}\right)
.\nonumber
\end{align}
Based on the charge operator $Q=\frac{1}{2} \lambda ^3 +
\frac{1}{2\sqrt{3}} \lambda ^8$ the $D_{Qa}^{(8)}$ is used as
abbreviation for
$D_{Qa}^{(8)}=\frac{1}{2}D_{3a}^{(8)}+\frac{1}{2\sqrt{3}}D_{8a}^{(8)}$.
The parameters $w_{1,2,3}$ are of order $\mathcal{O}(1)$, while
$w_{4,5,6}$ are of order $\mathcal{O}(m_s)$, $m_{s}$ being
regarded as a small parameter.

The full expression for the magnetic moments can be decomposed as follows:
\begin{equation}
\mu_{B}=\mu_{B}^{(0)}+\mu_{B}^{(op)}+\mu_{B}^{(wf)},
\end{equation}
where the $\mu_{B}^{(0)}$ is given by the matrix element of the
$\hat{\mu}^{(0)}$ between the purely symmetric states $\left|
\mathcal{R}_{J},B,J_{3}\right\rangle$, and the $\mu_{B}^{(op)}$ is
given as the matrix element of the $\hat{\mu}^{(1)}$ between the
symmetry states as well.  The wave function correction
$\mu_{B}^{(wf)}$ is given as a sum of the interference matrix
elements of the $\mu_{B}^{(0)}$ between purely symmetric states
and admixtures displayed in Eq.(\ref{admix}).  These matrix
elements were calculated for baryon octet and decuplet in
Ref.\cite{Kim:1998gt}.

Denoting the set of the model parameters by
\begin{equation}
\vec{w}=(w_{1},\ldots,w_{6})
\end{equation}
the formulae for the set of the magnetic moments in representation
$\mathcal{R}$ (of dimension $R$)%
\begin{equation}
\vec{\mu}^{\mathcal{R}}=(\mu_{B_{1}},\ldots,\mu_{B_{R}})
\end{equation}
can be conveniently cast into the form of the matrix equations:%
\begin{equation}
\vec{\mu}^{\mathcal{R}}=A^{\mathcal{R}}[\Sigma_{\pi N}]\cdot\vec{w},%
\end{equation}
where rectangular matrices $A^{8}$ and $A^{10}$ and
$A^{\overline{10}}$ (Note their dependence on the pion-nucleon
$\Sigma_{\pi N}$ term) can be found in
Refs.\cite{Kim:1997ip,Kim:1998gt,YangKim} in the basis

\begin{equation}
\vec{\mu}^{\overline{10}}=(\mu_{\Theta^{+}},\mu_{p^{\ast}},\mu_{n^{\ast}}%
,\mu_{\Sigma^{+}},\mu_{\Sigma^{0}},\mu_{\Sigma^{-}},\mu_{\Xi^{+}},\mu_{\Xi
^{0}},\mu_{\Xi^{-}},\mu_{\Xi^{--}}).
\end{equation}

In order to find the set of parameters $w_{i}[\Sigma_{\pi N}]$, we
minimize the mean square deviation for the octet magnetic moments:
\begin{equation}
\Delta\mu^{8}=\frac{1}{7}\sqrt{\sum_{B}\left(
    \mu_{B,\,th}^{8}[\Sigma_{\pi N}]-\mu_{B,\,exp}^{8}\right)  ^{2}},
\end{equation}
where the sum extends over all octet magnetic moments, but the
$\Sigma^{0}$.  The value $\Delta\mu^{8}\simeq0.01$ is in practice
independent of the $\Sigma_{\pi N}$ in the physically interesting range
$45$ $-$ $75$ MeV.  The values of the $\mu_{B,\,th}^{8}[\Sigma_{\pi
N}]$ are independent of $\Sigma_{\pi N}$.  Table~\ref{tab:1}
lists the results of the magnetic moments of the baryon octet.
\begin{table}[h]
  \centering
  \begin{tabular}[c]{|l|ccccccc|}\hline
& $p$ & $n$ & $\Lambda^{0}$ & $\Sigma^{+}$ & $\Sigma^{-}$ & $\Xi^{0}$
& $\Xi^{-}$ \\ \hline
th. & $2.814$ & $-1.901$ & $-0.592$ & $2.419$ & $-1.172$ & $-1.291$ & $-0.656$\\
exp. & $2.793$ & $-1.913$ & $-0.613$ & $2.458$ & $-1.16$~ &
$-1.25$~ & $-0.651$ \\ \hline
  \end{tabular}
  \caption{Magnetic moments of the baryon octet.}
\label{tab:1}
\end{table}

Similarly, the value of the nucleon strange magnetic moment is independent of
$\Sigma_{\pi N}$ and reads $\mu_{N}^{(s)}=0.39 \,{\rm n.m.}$ in fair
agreement with our previous analysis of Ref.\cite{Kim:1998gt}.
Parameters $w_{i}$, however, do depend on $\Sigma_{\pi N}$.  This is
shown in Table.\ref{tablewi}:
\begin{table}[h]
  \centering
\begin{tabular}
[c]{|c|cccccc|}\hline
\multicolumn{1}{|c|}{$\Sigma_{\pi N}$ [MeV]} & $w_{1}$ & $w_{2}$ &
$w_{3}$ &   $w_{4}$ & $w_{5}$ & $w_{6}$ \\ \hline
$45$ & $-8.564$ & $14.983$ & $7.574$ & $-10.024$ & $-3.742$ & $-2.443$\\
$60$ & $-10.174$ & $11.764$ & $7.574$ & $-9.359$ & $-3.742$ & $-2.443$\\
$75$ & $-11.783$ & $8.545$ & $7.574$ & $-6.440$ &  $-3.742$ & $-2.443$
\\ \hline
\end{tabular}
\caption{Dependence of the parameters $w_i$ on $\Sigma_{\pi N}$.}
\label{tablewi}
\end{table}
Note that parameters $w_{2,3}$ are formally $\mathcal{O}(1/N_{c})$
with respect to $w_{1}$. As for smaller $\Sigma_{\pi N}$, this
$N_{c}$ counting is not born by explicit fits.

The magnetic moments of the baryon decuplet and antidecuplet
depend on the $\Sigma_{\pi N}$.  However, the dependence of the
decuplet is very weak.  The results are summarized in Table
\ref{table10},
\begin{table}[h]
  \centering
\begin{tabular}
[c]{|c|cccccccccc|}\hline
$\Sigma_{\pi N}$ [MeV] & $\Delta^{++}$ & $\Delta^{+}$ & $\Delta^{0}$ &
$\Delta^{-}$ & $\Sigma^{\ast+}$ &
$\Sigma^{\ast0}$ & $\Sigma^{\ast-}$ & $\Xi^{\ast0}$ & $\Xi^{\ast-}$ &
$\Omega^{-}$\\\hline
$45$ & $5.40$ & $2.65$ & $-0.09$ & $-2.83$ & $2.82$ & $0.13$ & $-2.57$
& $0.34$ & $-2.31$ & $-2.05$\\
$60$ & $5.39$ & $2.66$ & $-0.08$ & $-2.82$ & $2.82$ & $0.13$ & $-2.56$
& $0.34$ & $-2.30$ & $-2.05$\\
$75$ & $5.39$ & $2.66$ & $-0.07$ & $-2.80$ & $2.81$ & $0.13$ & $-2.55$
& $0.33$ & $-2.30$ & $-2.05$\\
\text{Ref.\cite{Kim:1997ip}} & $5.34$ & $2.67$ & $-0.01$ & $-2.68$ &
$3.10$ & $0.32$ & $-2.47$ & $0.64$ & $-2.25$ & $-2.04$ \\ \hline
\end{tabular}
\caption{Magnetic moments of the baryon decuplet.}
\label{table10}%
\end{table}
where we also display the theoretical predictions from
Ref.\cite{{Kim:1997ip}} for $p=0.25$.  Let us note that the $m_{s}$
corrections are not large for the decuplet and the approximate
proportionality of the $\mu_{B}^{10}$ to the baryon charge
$Q_{B}$ still holds.

\begin{figure}[h]
\begin{center}
\includegraphics[scale=1.1]{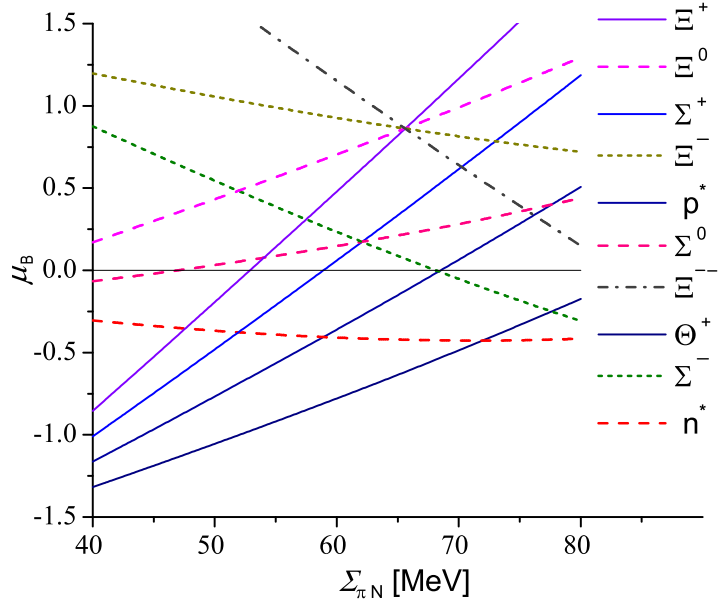}
\end{center}
\caption{Magnetic moments of antidecuplet as functions of $\Sigma_{\pi N}$.}%
\label{fig:10bar}%
\end{figure}

Finally,  for anti-decuplet we have a strong dependence on
$\Sigma_{\pi N}$, yielding the numbers of Table \ref{tab10b}.
\begin{table}[h]
  \centering
\begin{tabular}
[c]{|c|cccccccccc|}\hline
$\Sigma_{\pi N}$ [MeV] & $\Theta^{+}$ & $p^{\ast}$ & $n^{\ast}$ &
$\Sigma_{\overline{10}}^{+}$ & $\Sigma_{\overline{10}}^{0}$ &
$\Sigma_{\overline{10}}^{-}$ & $\Xi_{\overline{10}}^{+}$ &
$\Xi_{\overline{10}}^{0}$ & $\Xi_{\overline{10}}^{-}$ &
$\Xi_{\overline {10}}^{--}$ \\\hline
$45$ & $-1.19$ & $-0.97$ & $-0.34$ & $-0.75$ & $-0.02$ & $\;0.71$ &
$-0.53$ & $0.30$ & $1.13$ & $1.95$\\
$60$ & $-0.78$ & $-0.36$ & $-0.41$ & $\;0.06$ & $\;0.15$ & $\;0.23$ &
$\;0.48$ & $0.70$ & $0.93$ & $1.15$\\
$75$ & $-0.33$ & $\;0.28$ & $-0.43$ & $\;0.90$ & $\;0.36$ & $-0.19$ &
$\;1.51$ & $1.14$ & $0.77$ & $0.39$ \\ \hline
\end{tabular}
\caption{Magnetic moments of the baryon antidecuplet.}
\label{tab10b}%
\end{table}
The results listed in Table~\ref{tab10b} are further depicted in
Fig.\ref{fig:10bar}.

In the chiral limit, the antidecuplet magnetic moments are
proportional to the corresponding charges, but with opposite sign,
and they read numerically
\begin{equation}
\mu_{B}^{{\overline{10}}\;(0)}=-(1.05\sim 0.24)Q_{B}\label{10bchiral}%
\end{equation}
for $\Sigma_{\pi N}=45$ and $75$ MeV, respectively. The inclusion
of the $m_{s}$ corrections introduces splittings and
proportionality to the charge is violated.  The magnitude of the
splitting increases as the $\Sigma_{\pi N}$ does. As described in
Ref.~\cite{YangKim} for small $\Sigma_{\pi N}$ corrections due to
the nonzero $m_{s}$ are moderate and the perturbative approach is
reliable.  On the contrary, for large $\Sigma_{\pi N}$,
corrections are large.  It is due to the wave function
corrections, since the dependence of the operator part on the
$\Sigma_{\pi N}$ given in terms of the coefficients $w_{4,5,6}$ is
small as in Table (\ref{tablewi}). The wave function corrections
tend to be cancelled for the non-exotic baryons and add
constructively for the baryon antidecuplet.  In particular, for
$\Sigma_{\pi N}=75$~MeV we have a large admixture coefficient of
27-plet which tend to dominate the otherwise small magnetic
moments of anti-decuplet. At this point, the reliability of the
perturbative expansion for the antidecuplet magnetic moments may
be questioned. On the other hand, as remarked above, the $N_{c}$
counting for the $w_{i}$ coefficients works much better for large
$\Sigma_{\pi N}$. One notices for reasonable values of
$\Sigma_{\pi N}$ some interesting facts, which were partially
reported already in Ref.\cite{Kim:2003ay}: The magnetic moments of
the baryon antidecuplet are rather small in absolute value. For
$\Theta^+$ and $p^*$ one obtains negative values although the
charges are positive.  For $\Xi^{-}_{\overline{10}}$ and
$\Xi^{--}_{\overline{10}}$ one obtains positive values although
the signs of the charges are negative.

Actually, the magnetic moments of the positive-parity pentaquarks
have been studied by a number of authors in different models
\cite{Zhao:2003gs,Huang:2003bu,Liu:2003ab,Bijker:2004gr,Hong:2004xn}.
The results are displayed in Table~\ref{table5}.  We see that in
all quark models the magnetic moment of the $\Theta^+$ is rather
small and positive.  On the contrary, our present model-independent
analysis shows that $\mu_{\Theta^+}< 0$, although the
magnitude depends strongly on the value of $\Sigma_{\pi N}$.  The
measurement of $\mu_{\Theta^+}$ could therefore discriminate
between different models. This also may add to reduce the
ambiguities in the pion-nucleon sigma term $\Sigma_{\pi N}$.

The measurement of the antidecuplet magnetic moments by ordinary
precession techniques is not possible.  However, it is crucial to
know the magnetic moment of the $\Theta^{+}$ in order to study its
production via photo-reactions. One can use the measured cross
section to determine the magnetic moment of the $\Theta^{+}$.  The
cross sections for the $\Theta^{+}$ production from nucleons
induced by photons~\cite{Nametal} have been already described
theoretically.  A similar approach was used to determine the
magnetic moments of the $\Delta^{++}$~\cite{Deltapp1,Deltapp2} and
$\Delta^{+}$~\cite{Kotulla:2002cg}, which are much broader than
the $\Theta^{+}$.  The measurements of the $\Delta^{++}$ magnetic
moment comes from the reaction such as
$\pi^{+}p\rightarrow\pi^{+}\gamma^{\prime}p$~\cite{Deltapp1,Deltapp1},
while that of the $\Delta^{+}$ was measured in $\gamma
p\rightarrow\pi^{0}\gamma^{\prime}p$~\cite{Kotulla:2002cg}. This
shows that the measurement of the magnetic moments of resonances
is in principle possible, despite the fact that it is difficult
and is hampered by large uncertainties which mainly come from the
systematic error of cross-section calculations.

\begin{table}
  \centering
\label{table5}
\begin{tabular}[c]{|cl|lcccc|}\hline
Ref. & first author & model & remarks & ~$\Theta^{+}$ & ~$\Xi^{--}$ & ~$\Xi^{+}%
$\\ \hline
\cite{Zhao:2003gs} & Q. Zhao & diquarks (JW) &  &  $~~0.08$ &  $-$ &  $-$\\
\hline
\cite{Huang:2003bu}& P.Z. Huang & sum rules & abs. value &  $0.12\pm0.06$ &  $-$ &  $-$\\
\hline
\cite{Liu:2003ab}& Y.-R. Liu & diquarks (JW) &  &  $~~0.08$ &  $~~0.12$ &  $-0.06$\\
&  & clusters (SZ) &  &  $~~0.23$ &  $-0.17$ &  $~~0.33$\\
&  & triquarks (KL) &  &  $~~0.37$ &  $~~0.43$ &  $~~0.13$\\
&  & MIT bag (S) &  &  $~~0.37$ &  $-0.42$ &  $~~0.45$\\
\hline
\cite{Bijker:2004gr}& R. Bijker & QM, harm.osc. &  &  $~~0.38$ &  $-0.44$ &  $~~0.50$\\
\hline
\cite{Hong:2004xn}& D.K. Hong & chiral eff. th. &  $m_{s}=400$ MeV &  $~~0.71$ &  $-$ &  $-$\\
&  & with diquarks &  $m_{s}=450$ MeV &  $~~0.56$ &  $-$ &  $-$\\
\hline
\multicolumn{2}{|c|}{present  work}& chiral soliton &  $\Sigma_{\pi N}=45$ MeV &  $-1.19$ &
$~~1.95$ &  $-0.53$\\
&  & model & $\Sigma_{\pi N}=75$ MeV &  $-0.33$ &  $~~0.39$ &
$~~1.51$ \\ \hline
\end{tabular}
\caption{Magnetic moments of $\Theta^+$, $\Xi^{--}$ and $\Xi^{+}$ in nuclear
magnetons from different papers in different models. (JW) stands for Jaffe
and Wilczek \cite{Jaffe:2003sg}, (SZ) for Shuryak and Zahed \cite{Shuryak:2003zi},
(KL) for Karliner and Lipkin \cite{Karliner:2003dt} and (S) for Strottman
\cite{Strottman:1979qu}.}
\end{table}

%

\section{Conclusion and Summary}

The present paper can be summarized as follows:

The estimates in the framework of the chiral soliton were prior to
the experiments and their predictions concerning mass and width
have so far been confirmed by several measurements. All other
calculations were produced after the first experiments.

The solitonic approach relies in its theoretical background on the
limit of large $N_c$ in order to justify the stationary phase
approximation and the adiabatic rigid-rotor quantization. While
this is no problem for the baryon octet and decuplet the
description of the exotic $\Theta ^+$ yields a $N_c$-dependence
which results in a possible mixing with vibrations. Thus the
predictions of the solitonic approach need further dynamical
arguments for its justification, which are yet unknown. The
solitonic approach provides a intuitive explanation for the
smallness of the width of $\Theta ^+$ by the overlap of the five-quark
component of the light-cone wave function of the nucleon
with the five quark component of $\Theta ^+$. This picture is
based fully on the Goldstone character of the kaon field.

The most popular of the quark models is the one by Jaffe and
Wilczek. Although they do not calculate the width of $\Theta ^+$
they have clear predictions for several excited baryonic states. In
particular they predict nearly degenerate antidecuplet of spin 3/2
which does not exist in the chiral models. This feature may be
used to discriminate between the two. There are criticisms based
on the model of the instanton liquid of the vacuum that their diquarks
should provide higher mass of the $\Theta ^+$. There is also
criticism that their predictions on nucleon excited states do not
match the presently known phenomenology if one uses the model
calculations of Jaffe and Wilczek and uses in addition the fact
that experiments yield a very small width of $\Theta ^+$.

Lattice gauge calculations for $\Theta ^+$ are presently hampered
for various reasons: The quark sources are not properly known for,
those, which are used, are oversimplified. The mass of the $\Theta
^+$ is 100 MeV above the kaon-nucleon threshold and hence, in a
discretized calculation, the separation of the bound state and the
scattering states is difficult. Furthermore the extrapolation from
large (technical) pion masses to physical ones is not well under
control.

In summary: The chiral soliton model has been extremely successful
in predicting for the first time an exotic baryon and its relevant
properties. However, the physics behind this prediction is not yet
fully understood in particular with respect to the large-$N_c$
limit.\\

Still the soliton model seems to provide the most promising
approach. Hence it is useful to apply it to further properties of
the $\Theta ^+$ and antidecuplet.
 This is done for the magnetic moments, using the so-called
model-independent approach based on rigid-rotor quantization
including strange mass corrections. This yields
magnetic moments for the antidecuplet which are small in absolute
size and of opposite sign compared to the charge. The value
depends strongly on the value of the pion-nucleon sigma term and
is in the range $-1.19\,{\rm n.m.}$ to $-0.33\,{\rm n.m.}$. As a
byproduct the strange magnetic moment of the nucleon can be
evaluated to $\mu^{(s)}_N =+0.39$n.m.

\section*{Acknowledgments}
M.P. acknowledges the hospitality of the Nuclear Theory Group at
Brookhaven National Laboratoty (BNL) where parts of this work have
been completed. The present work is supported by Korea Research
Foundation Grant: KRF-2003-041-C20067 (H.-Ch.K.) and by the Polish
State Committee for Scientific Research under grant 2 P03B 043 24
(M.P.) and by Korean-German (F01-2004-000-00102-0) and
Polish-German grants of the Deutsche Forschungsgemeinschaft (DFG).
The work is partially supported by the
Transregio-Sonderforschungsbereich Bonn-Bochum-Giessen as well as
by the Verbundforschung and the International Office of the
Federal Ministry for Education and Research (BMBF). Fruitful
discussions with Tom Cohen, Maxim Polyakov, Pavel Pobylitsa,
Antonio Silva, Fl. Stancu, V. Guzey and Peter Schweitzer are
acknowledged.

\end{document}